\documentclass[epj]{svjour}
\usepackage{hyperref}
\usepackage[cp1250]{inputenc}
\usepackage{amsmath}
\usepackage{amssymb}
\usepackage{graphicx}
\usepackage{url}
\usepackage{booktabs}
\usepackage{array}
\begin{document}
\title{Information-theoretic approach to lead-lag effect on financial markets}
\author{Paweł Fiedor}
\institute{Cracow University of Economics, Rakowicka 27, 31-510 Kraków, Poland, \email{s801dok@wizard.uek.krakow.pl}}
\date{\today}
\abstract{Recently the interest of researchers has shifted from the analysis of synchronous relationships of financial instruments to the analysis of more meaningful asynchronous relationships. Both of those analyses are concentrated only on Pearson's correlation coefficient and thus intraday lead-lag relationships associated with such. Under Efficient Market Hypothesis such relationships are not possible as all information is embedded in the prices. In this paper we analyse lead-lag relationships of financial instruments and extend known methodology by using mutual information instead of Pearson's correlation coefficient, which not only is a more general measure, sensitive to non-linear dependencies, but also can lead to a simpler procedure of statistical validation of links between financial instruments. We analyse lagged relationships using NYSE 100 data not only on intraday level but also for daily stock returns, which has usually been ignored.
}
\maketitle

\section{Introduction}

Financial markets are becoming increasingly more complex as adaptive systems. Nonetheless economists did lack a fundamental theory behind their complex behaviour even at times when their structure has been much simpler. This lack of theory has many consequences. First, other scientists, notably physicists, can study those systems without worrying about the intricacies of economic theory. Second, the lack of theory leads to an assumption that the time series describing stock returns are unpredictable \cite{Samuelson:1965}. Within this paradigm the evolution of stock prices can only be explained by random processes. Additionally the Efficient Market Hypothesis \cite{Tobin:1969} proposes that all information is reflected in the prices and that it is not possible to predict future prices based on the past. There are weaker variants of the hypothesis stating that only past prices are included in the current ones, thus rendering predictions based on the past prices only impossible. This hypothesis would then mean that there can be no lead-lag effect on the financial markets, making the analysis in the paper pointless. But the Efficient Market Hypothesis has been continually disproved in many ways since the 1980s, and in fact the support for it has dwindled among researchers. Particularly researchers analysing NYSE stock returns \cite{Lo:1988,Shmilovici:2003} show that the data can be compressed, thus showing that the stock returns are not random, as then no compression would be possible. We have performed similar tests on New York and Warsaw exchanges in the recent past \cite{Fiedor:2014}. Then if the price changes of stocks are not random there arises a possibility that the data is structured. Thus researchers are encouraged to explore methods of modelling this structure and analysing real-world markets.

The assumption that price formation are stochastic processes leaves researchers with a question whether these processes are independent for different financial instruments or whether there exist relationships based on known or unknown common economic factors driving these formation processes. Tools and procedures developed first to model physical systems \cite{Mandelbrot:1963,Kadanoff:1971,Mantegna:1991} are often used to analyse the interdependencies between financial instruments. The most effort has been used in understanding correlations in financial markets for daily \cite{Mantegna:1999,Cizeau:2001,Forbes:2002,Podobnik:2008,Aste:2010,Kenett:2012meta} and intraday time scales \cite{Bonanno:2001,Tumminello:2007,Munnix:2010}. In the recent years other measures of similarity have also been introduced, including Granger-causality analysis \cite{Billio:2002}, partial correlation analysis \cite{Kenett:2010}, both of which try to quantify how one financial instrument provides information about another instrument, and mutual information \cite{Fiedor:2014a} together with mutual information rate \cite{Fiedor:2014b}, both of which aim at including non-linear relationships in the analysis. All of these methods aim for a single goal, that is the discovery of meaningful information in the increasingly complex adaptive systems of financial markets.

The most common analysis uses synchronous correlations of equity returns. Such analyses have shown that financial markets have a nested structure in which stock returns are driven by a common factor, and stocks themselves are organised in groups by sector. The correlations inside those groups are higher than the average pair correlation. One can also find second order groups, that is within the sectors one can distinguish groups of stocks belonging to the same sub-sector, which display an even higher correlation. The correlations can of course be exchanged for another well-defined similarity measure such as mutual information \cite{Fiedor:2014b}. This is well corroborated as the same results have been obtained using substantially different methods, ranging from random matrix theory \cite{Laloux:2000}, through principal component analysis \cite{Fenn:2011}, through hierarchical clustering \cite{Mantegna:1999}, to correlation-based networks \cite{Mantegna:1999,Bonanno:2003,Onnela:2003} and mutual information-based networks \cite{Fiedor:2014a}. The methods developed to construct dependency networks may be grouped into two categories: threshold-based methods and topological methods. Both categories start with a sample similarity measure (correlation matrix, mutual information matrix etc.). Then using the threshold method a threshold is set on the similarity measure and a network is constructed in which only links between nodes whose pairwise similarity measure is larger than the threshold are present. With lowering the threshold value a more complex hierarchy emerges, and a formation of groups of stocks progressively merge to form larger groups, until they form the whole market. Such threshold networks are very robust with regards to the uncertainty in the similarity measure, but it is difficult to find a single threshold value which could accurately display the nested structure of the similarity matrix of stock returns. Topological methods on the other hand construct dependency networks, such as the minimal spanning tree (MST) \cite{Mantegna:1999,Bonanno:2003,Onnela:2003,Fiedor:2014a} or the planar maximally-filtered graph (PMFG) \cite{Tumminello:2005,Tumminello:2010,Fiedor:2014a}, are based on the ranking of empirical similarity measures. The resulting networks are intrinsically hierarchical and therefore easy to be presented as a graph, but this approach is less stable than threshold methods with respect to the statistical uncertainty in the data. Futhermore such approach does not necessarily present information about the statistical significance of the similarity measures \cite{Coronnello:2007}.

On the other hand very few inquires have been performed looking into networks of lagged correlations \cite{Huth:2011,Curme:2014}. The above-described methods of constructing dependency networks cannot be easily extended to the analysis of directed lagged correlations or similarity measures in financial markets. The lagged interdependencies in stock returns are quite small even at short time horizons, therefore an analysis is strongly influenced by the statistical uncertainty of the estimation process. The use of topological methods is difficult as they only take into consideration the ranking of similarity measures and not their actual values, thus many links in such a network may indeed be statistically insignificant if we use lagged dependencies. On the other hand, threshold methods are difficult to apply because it is difficult to find an appropriate threshold level. Also these methods use the same threshold for all stock pairs, which is a problem in the analysis of lagged relationships, as the statistical significance of a lagged similarity measure is likely to vary across stocks (for example due to different volatility).

In \cite{Curme:2014} a method for filtering a lagged correlation matrix into a network of statistically-validated directed links that takes into account the heterogeneity of stock return distributions has been introduced. This has been done by associating a $p$-value with each observed lagged-correlation and then setting a threshold on $p$-values, i.e., setting a level of statistical significance corrected for multiple hypothesis testing. They have applied this method to analyse the structure of lagged relationships between intraday equity returns on US equity markets.

In this paper we are extending this analysis in two ways. First, we extend this methodology to include non-linear relationships. Second, we also analyse daily lagged relationships. It is well-known that financial markets, and particularly time series describing returns on financial instruments, are involving terms that are not of the first degree. There is now strong evidence of the existence of non-linear dynamics in stock returns \cite{Brock:1991,Qi:1999,McMillan:2001,Sornette:2002,Kim:2002}, market index returns \cite{Franses:1996,Wong:1995,Chen:1996,Wong:1997,Ammermann:2003}, and currency exchange rate changes \cite{Hsieh:1989,Brock:1991,Rose:1991,Brooks:1996,Wu:2003}. 

Meanwhile Pearson's correlation coefficient is strictly not sensitive to any non-linear dependencies. Therefore an analysis using correlation can miss important features of any dynamical system, particularly financial markets. Thus we find the assumptions that only linear dependencies are relevant in financial markets found in hierarchical clustering methodology used in econophysics unsupportable. We contrast correlation coefficient is then contrasted by the measure of mutual information ($I_S$) \cite{Cover:1991}, which is a more general measure. In fact $I_S = 0$ if and only if the two studied random variable are strictly independent. Mutual information is a natural measure which can be used to extend the similarity measure to make it sensitive to non-linear dependencies, and has been successfully used in some applications \cite{Zhou:2007,Zhou:2009,Muller:2012}. Recently we have used it in the creation of dependency networks on financial markets \cite{Fiedor:2014b}. Mutual information is a measure of great importance in many fields precisely because it quantifies both the linear and non-linear interdependencies between two stochastic processes. Mutual information measures how much information two studied stochastic processes share. Mutual information is suitable for many applications and has been used to enhance the understanding of the brain in neuroscience \cite{Sporns:2004,Bialek:1998,Bialek:2002}, to characterise \cite{Donges:2009,Palus:2001} and model various complex and chaotic systems \cite{Fraser:1986,Parlitz:1998,Kantz:2004}, and also to quantify the information capacity of a communication system \cite{Haykin:2001}. Additionally mutual information provides a convenient way to identify the most relevant variables with which to describe the behaviour of a complex system \cite{Rossi:2006}, which is of paramount importance in modelling those systems, and indeed to the methodology of this paper \cite{Fiedor:2014a,Fiedor:2014b}.

Furthermore we have found in our earlier studies \cite{Fiedor:2014} that while intraday stock returns are deviating from EMH much stronger than daily returns, the latter themselves are not random, thus we will also look into lead-lag relationships in the daily stock returns. We believe that lead-lag effect will be much smaller in daily stock returns, but nonetheless it may not be negligible.

The paper is organised as follows. In Sect.~2 we present the method used to filter and validate statistically significant lagged correlations and introduce a method for statistical validation of significant mutual information between financial instruments. In Sect.~3 we analyse the structure of NYSE at different frequencies using the presented methodology. In Sect.~4 we discuss the results. In Sect.~5 we conclude the study.

\section{Methods}

Here we present the methodology of statistically validating lagged correlations for the purpose of network analysis presented in \cite{Curme:2014}. On this basis we will present our extended methodology which includes non-linear dependencies. For this purpose we will also need to define mutual information, its properties and estimators.

Curme et al. \cite{Curme:2014} begin the analysis by calculating the matrix of logarithmic returns over given intraday time-horizons. Let us denote the most recent price for stock $n$ occurring on or before time $t$ during the trading day by $p_n(t)$. The opening price of the stock is defined as the price of its first transaction of the trading day. Additionally $\tau$ is the time horizon. Then for each stock the logarithmic returns are sampled,
\begin{equation}
 r_{n,t} \equiv \log(p_n(t)) - \log(p_n(t-\tau)),
\end{equation}
every $\tau$ minutes (days, seconds) throughout the trading day. These time series constitute columns in a matrix $R$. Then $R$ is filtered into two matrices, $A$ and $B$, in which returns during the last period $\tau$ of each trading day are excluded from $A$ and returns during the first period $\tau$ of each trading day are excluded from $B$. From these matrices an empirical lagged correlation matrix $C$ is constructed using the Pearson correlation coefficient of columns of $A$ and $B$,
\begin{equation}
C_{m,n} = \dfrac{1}{T-1} \sum_{i=1}^T\dfrac{(A_{m,i}-\langle A_m\rangle) (B_{n,i}-\langle B_n\rangle)}{\sigma_m\sigma_n},
\label{eqn:corr_matrix}
\end{equation}
where $\langle A_m\rangle$ and $\sigma_m$ are the mean and sample standard deviation, respectively, of column $m$ of $A$, and $T$ is the number of rows in $A$ (and $B$). Curme et al. \cite{Curme:2014} set the lag $\lambda$ to be one time horizon $\tau$.

The matrix $C$ can be seen as a weighted adjacency matrix for a fully connected, directed graph. Such matrix needs to be filtered, and to find a threshold of statistical significance Curme et al. \cite{Curme:2014} apply a shuffling technique \cite{Efron:1993}. The rows of $A$ are shuffled repeatedly without replacement in order to create a large number of surrogate time series of returns. After each shuffling the lagged correlation matrix is recalculated as $\widetilde{C}$ and compared to the empirical matrix $C$. For each shuffling there is an independent realisation of $\widetilde{C}$. Then matrices $U$ and $D$ are constructed, where $U_{m,n}$ is the number of realisations for which $\widetilde{C}_{m,n} \geq C_{m,n}$, and $D_{m,n}$ is the number of realisations for which $\widetilde{C}_{m,n} \leq C_{m,n}$.

From matrix $U$ a one-tailed $p$-value is associated with all positive correlations as the probability of observing a correlation that is equal to or higher than the empirically-measured correlation. Similarly, from $D$ a one-tailed $p$-value is associated with all negative correlations. Curme et al. \cite{Curme:2014} set the threshold at the standard $p=0.01$. The statistical threshold must be adjusted to account for multiple comparisons. Curme et al. \cite{Curme:2014} use the conservative Bonferroni correction and a less conservative FRD adjustment which both depend on the sample size of $N$ stocks. In particular Bonferroni correction works as follows: $p/N^2$. For $N=100$ it gives $0.01/100^2$, thus in such case a construction of $10^6$ independently shuffled surrogate time series is required. If $U_{m,n}=0$ then a statistically-validated positive link from stock $m$ to stock $n$ ($p=0.01$, Bonferroni correction) can be associated. Likewise, if $D_{m,n}=0$ a statistically-validated negative link from stock $m$ to stock $n$ is associated. In this way Curme et al. \cite{Curme:2014} construct the Bonferroni network \cite{Tumminello:2011}.

Curme et al. \cite{Curme:2014} also construct an FDR network, using $p$-values corrected according to the false discovery rate (FDR) protocol \cite{Benjamini:1995}. This correction is less conservative than the Bonferroni correction. The $p$-values from each individual test are arranged in increasing order ($p_1 < p_2 < \dots < p_{N^2}$), and the threshold is defined as the largest $k$ such that $p_k < k~0.01/N^2$. In the FDR network the threshold for the matrices $U$ or $D$ is therefore not zero but the largest integer $k$ such that $U$ or $D$ has exactly $k$ entries fewer than or equal to $k$. From this threshold the links in $C$ can be filtered to construct the FDR network \cite{Tumminello:2011}. The Bonferroni network is a subgraph of the FDR network. This method makes no assumptions about the return distributions, and also imposes no topological constraints on the Bonferroni or FDR networks \cite{Curme:2014}.

Since this method only analyses strictly linear relationships we define mutual information to use instead of Pearson correlation coefficient. To extend such measure to include non-linear dependencies we propose to base the topological arrangement of the nodes in a network on the mutual information. Mutual information is most often defined in the context of Shannon's entropy \cite{Shannon:1948}, which is a measure of uncertainty of a random variable $X$:
\begin{equation}
	\label{eq:Def_entropy}
H(X) = -\sum_{i} p(x_i) \log_2 p(x_i) 
\end{equation}
summed over all possible outcomes $\{x_i\}$ with respective probabilities of $p(x_i)$. Joint $(X,Y)$ and conditional $H(X|Y)$ entropies are also defined for two variables.

We can also define mutual information in Shannon's sense \cite{Shannon:1948}. For two discrete random variables $X$ and $Y$ mutual information between them is defined as:
\begin{equation}
I_S(X,Y) = \sum_{y\in{}Y}\sum_{x\in{}X}p(x,y)\log{\frac{p(x,y)}{p(x)p(y)}},
\end{equation}
where $p(x,y)$ is the joint probability distribution function of $X$ and $Y$ and $p(x)$ and $p(y)$ are the marginal probability distributions. For continuous variables the definition is analogous using probability density functions. Equivalently using entropy mutual information is defined as:
\begin{equation}
I_S(X,Y) = H(X) + H(Y) - H(X,Y).
\label{IS}
\end{equation}
Mutual information measures information shared between the two variables, therefore both linear and non-linear dependencies, hence using it to describe dependencies on financial markets seems natural. Mutual information is non-negative and $I_S(X,X)=H(X)$. We also note that for easy estimation we need discrete data, while stock returns are not discrete, thus we need to discretize them. For discussion of this step see below and \cite{Fiedor:2014,Fiedor:2014a,Fiedor:2014b}.

We also need an estimator of entropy for practical purposes. There is a large number of estimators and a presentation of these can be found in \cite{Beirlant:1997,Darbellay:1999,Paninski:2003,Daub:2004,Nemenman:2004}. In this study we will use the plug-in estimator of entropy and mutual information, as we want our analysis to be conservative (for the same reason we will be using the Bonferroni correction). Such estimator is the entropy of the empirical distribution \cite{Paninski:2003}:
\begin{equation}
\hat{H}_{emp}(X)=-\sum_{x\in{}X}\frac{\Lambda(x)}{n}\log{\frac{\Lambda(x)}{n}},
\end{equation}
where $\Lambda(x)$ is the number of data points having value $x$, and $n$ is the sample size. Such entropy estimators are consistently biased downward (hence conservative).

Based on such definition we proceed with the method presented in \cite{Curme:2014} only exchanging correlation coefficient with mutual information. Since mutual information doesn't distinguish between positive and negative relationships we do not need both $U$ and $D$ and can settle with $U$. This is not a problem as in this analysis the direction of the relationship is not particularly important and can be easily found anyway. Similar analyses have been performed outside of economic systems \cite{Francois:2006}. Nonetheless a less computationally expensive method can be presented, without introducing very strong assumptions. It has been shown that the mutual information between independent random variables ($X$ \& $Y$) when estimated from relative frequencies follows a very good approximation of Gamma distribution with parameters $\alpha=(|X|-1)(|Y|-1)/2$ and $\beta=1/(N\ln{}2)$ \cite{Goebel:2005,Dawy:2006}:
\begin{equation}
I_S(X,Y)\sim\Gamma(\frac{1}{2}(|X|-1)(|Y|-1),\frac{1}{N\ln{}2}),
\end{equation}
where $N$ is the sample size and $|X|$ and $|Y|$ denote the numbers of realizations of the random variables $X$ and $Y$.

Here we briefly explain why that's the case. Using the natural logarithm in entropy expression we can expand the expression for mutual information $I_S(X,Y)$ into a Taylor series about expansion point $p_{XY}\equiv{}p_X p_Y$ and obtain:
\begin{equation}
I_S(X,Y)\approx\frac{1}{2}\sum_{x}\sum_{y}\frac{(p(x,y)-p(x)p(y))^2}{p(x)p(y)}.
\end{equation}
This expression relates to the $\chi^2$ test with the same constant factor of $2N$. The direct proof that the above has a Gamma distribution is rather technical and will not be presented. However, the same fact can be easily derived from knowing the $\chi^2$ test variable follows a $\chi^2$ distribution (given the null hypothesis is true). Since $I_S=(\chi^2)/(2N\ln{}2)$, we can scale the $\chi^2$ distribution by the factor $2N\ln{}2$ and obtain a Gamma distribution \cite{Goebel:2005,Dawy:2006}.

Therefore to determine the significance of $I(A_m,B_n)$ from a sample study of length $N$ at a significance level $p$, we check the condition:
\begin{equation}
I_S(A_m,B_n)\geq\Gamma_{1-p}(\frac{1}{2}(|A_m|-1)(|B_n|-1),\frac{1}{N\ln{}2},
\end{equation}
where $\Gamma_{1-p}(\alpha,\beta)$ denotes the $(1-p)$-quantile of the Gamma distribution. This is sound as under null hypothesis $A_m$ and $B_n$ are independent. As in \cite{Curme:2014} we need to adjust $p$ using Bonferroni or FDR correction. We will use this method instead of shuffling, as the latter has already been analysed in \cite{Curme:2014}. Both methods should give reasonably similar results.

\section{Materials and Results}

To find mutual information-based lagged relationships in practise we have taken log returns for 98 securities out of 100 which constitute the NYSE 100, excluding two with incomplete data. These log returns are intraday (1-minute intervals). The data covers 15 days between the 21st October 2013 and the 8th of November 2013. The choice of data length as much smaller than what \cite{Curme:2014} have used is explained in two ways. First, for empirical applications it is often required to see fast dynamics and not dynamics evolving over decades. Second, the choice of data spanning over many years would raise questions about the homogeneousness of the studied sample. Additionally we note that our dataset has length of over 3000, which is sufficient. To analyse daily relationships we also look at the daily price time series of 91 securities traded on New York Stock Exchange (NYSE100) (the 9 missing stocks were excluded due to missing data). The data has been downloaded from Google Finance database available at http://www.google.com/finance/ and was up to date as of the 11th of November 2013, going 10 years back. The data is transformed in the standard way for analysing price movements, that is so that the data points are the log ratios between consecutive daily closing prices, as defined above, and those data points are, for the purpose of estimating mutual information, discretized into 4 distinct states. The states represent equal parts, therefore each state is assigned the same number of data points. This design means that the model has no unnecessary parameters and proved to be very efficient \cite{Steuer:2001,Navet:2008,Fiedor:2014}. The choice of quartiles is largely irrelevant (equivalently one can choose 8 or 16 bins), see the discussion in \cite{Fiedor:2014}.

\begin{figure}[tbh]
\centering
\includegraphics[width=0.35\textwidth]{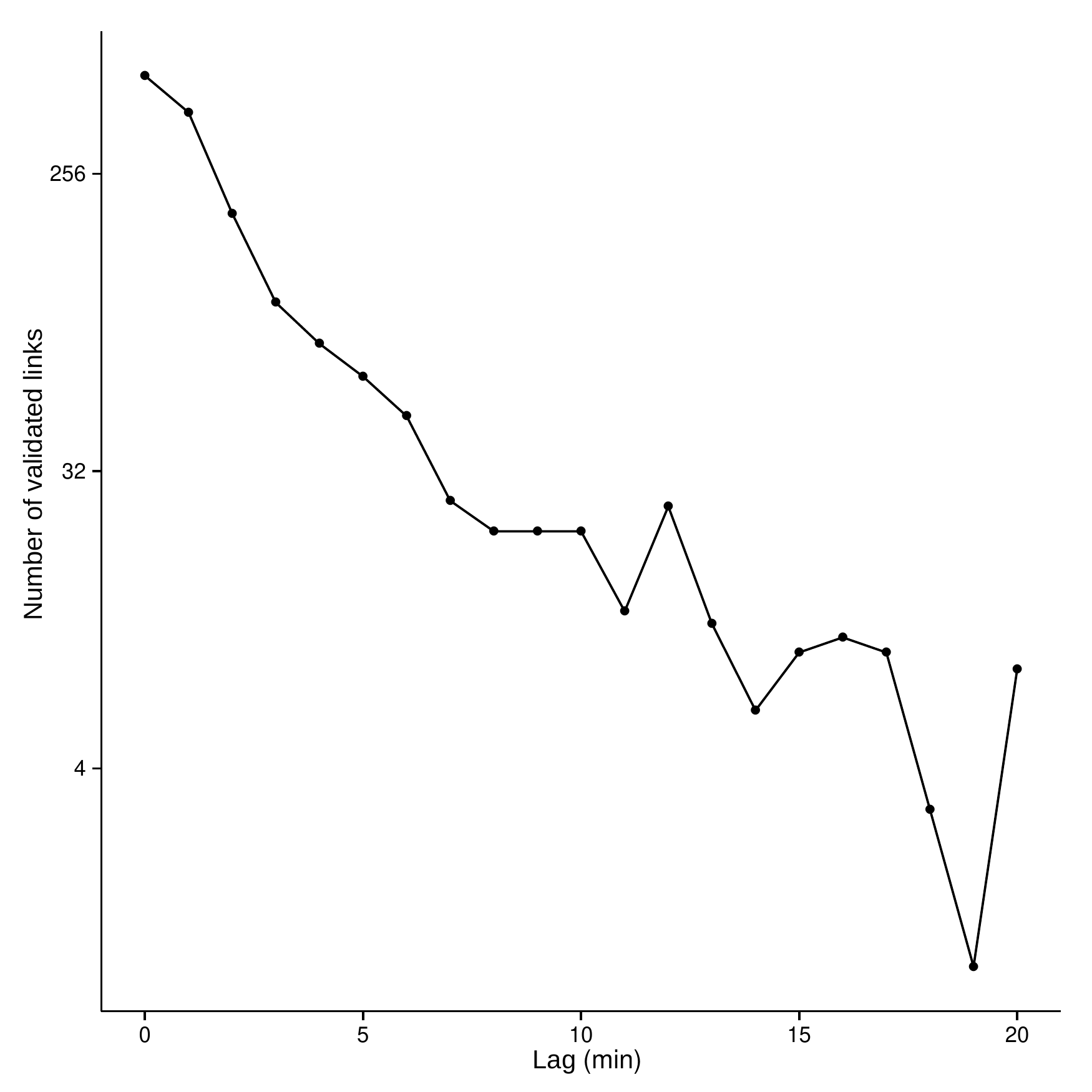}
\caption{Numer of validated links vs lag (intraday)}
\label{fig:nysemigamma}
\end{figure}

\begin{figure}[tbh]
\centering
\includegraphics[width=0.5\textwidth]{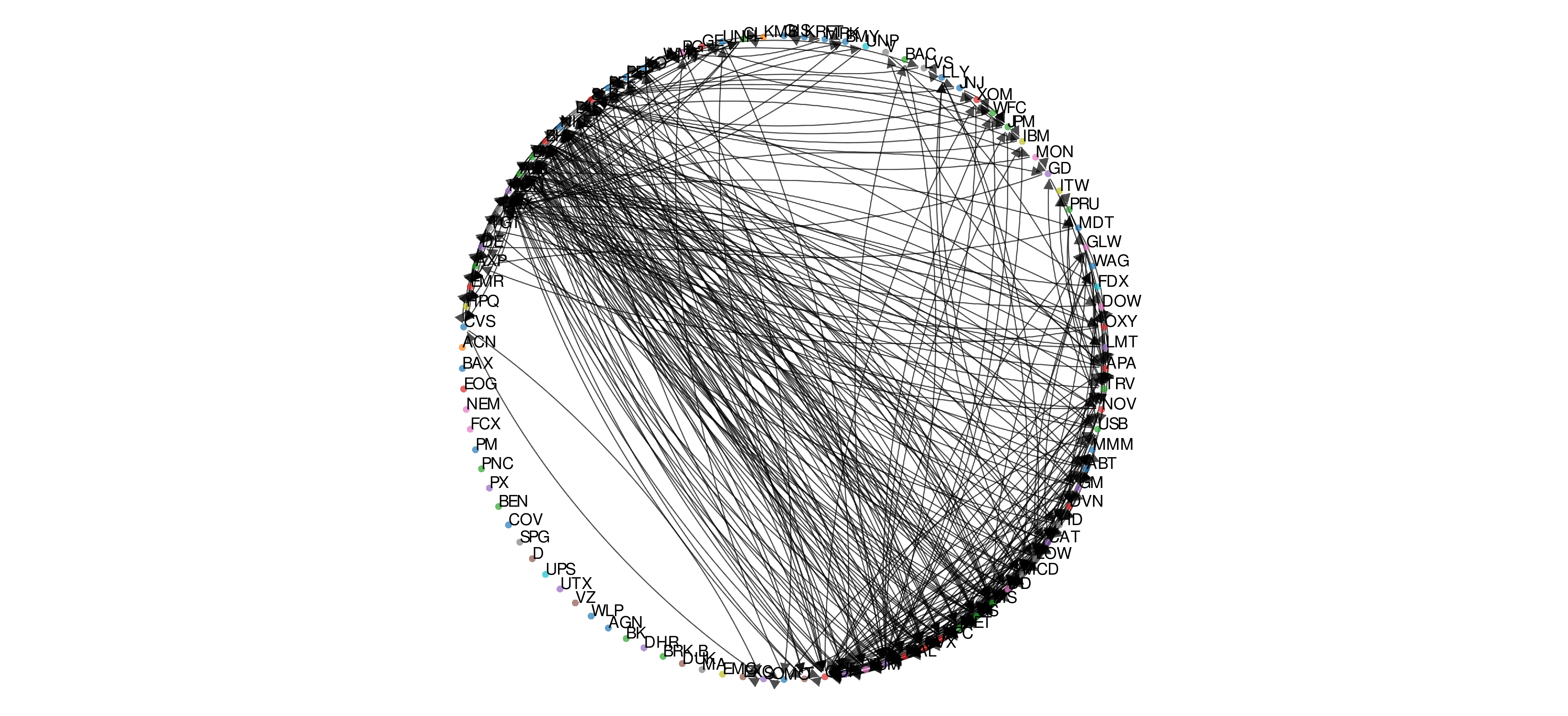}
\caption{MI-based network (intraday, $\lambda=0$)}
\label{fig:nyseid00}
\end{figure}

\begin{figure}[tbh]
\centering
\includegraphics[width=0.5\textwidth]{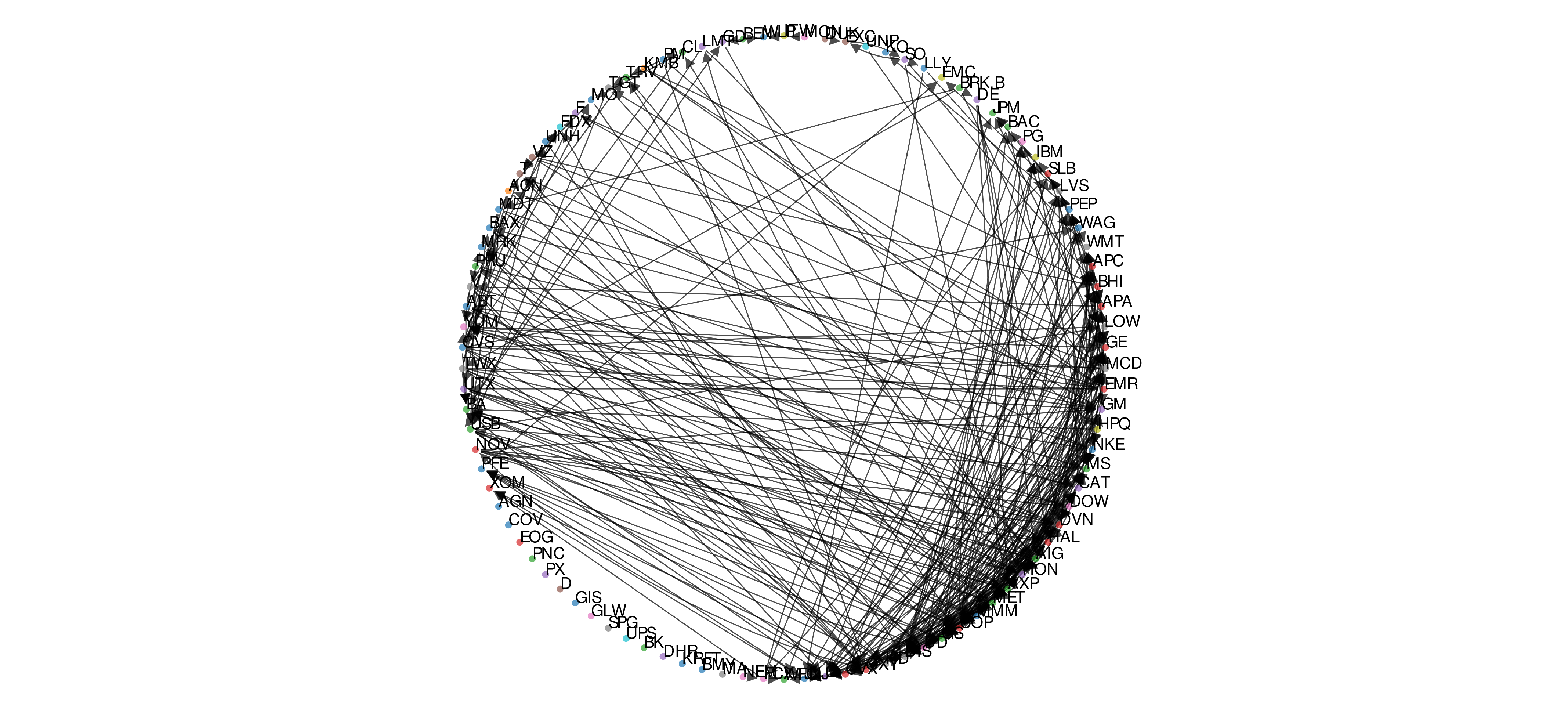}
\caption{MI-based network (intraday, $\lambda=1$)}
\label{fig:nyseid01}
\end{figure}

\begin{figure}[tbh]
\centering
\includegraphics[width=0.5\textwidth]{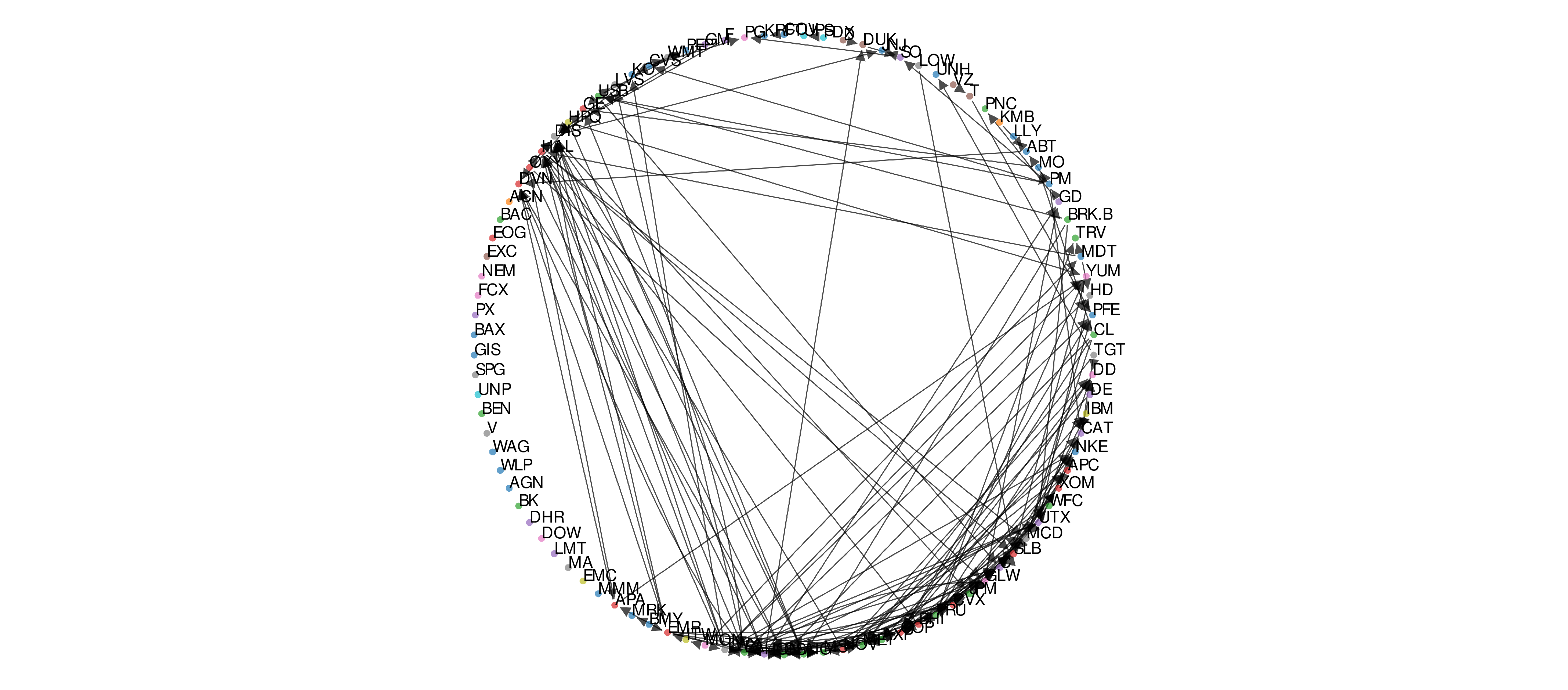}
\caption{MI-based network (intraday, $\lambda=2$)}
\label{fig:nyseid02}
\end{figure}

\begin{figure}[tbh]
\centering
\includegraphics[width=0.5\textwidth]{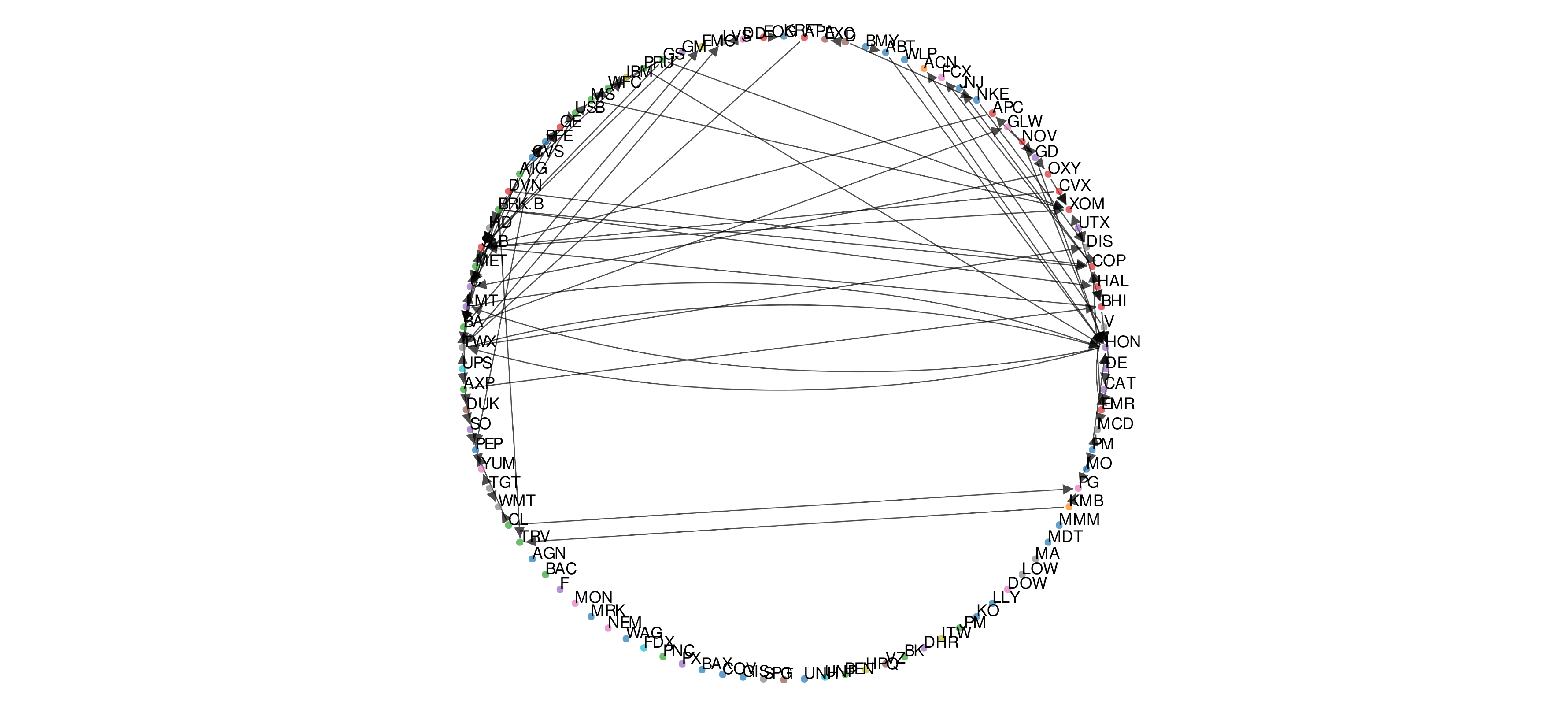}
\caption{MI-based network (intraday, $\lambda=3$)}
\label{fig:nyseid03}
\end{figure}

\begin{figure}[tbh]
\centering
\includegraphics[width=0.5\textwidth]{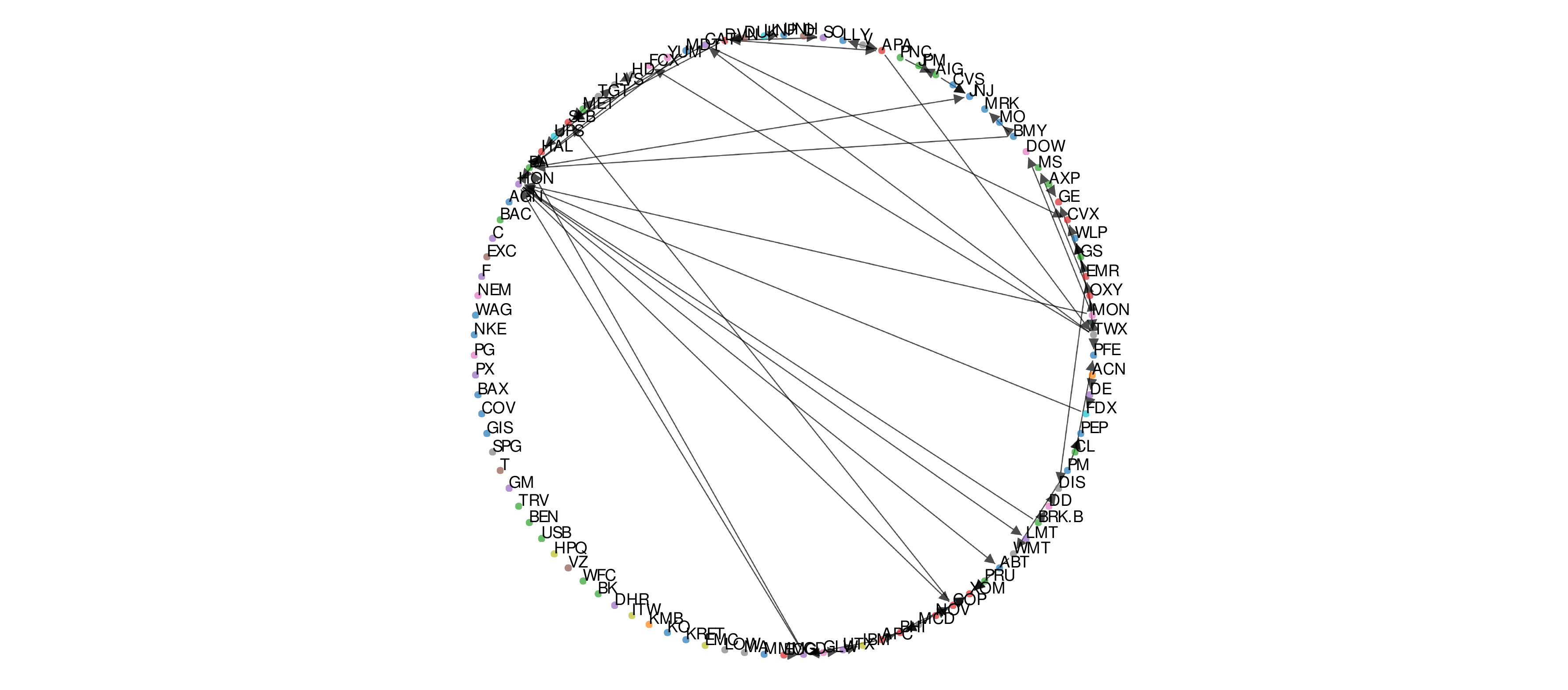}
\caption{MI-based network (intraday, $\lambda=4$)}
\label{fig:nyseid04}
\end{figure}

\begin{figure}[tbh]
\centering
\includegraphics[width=0.5\textwidth]{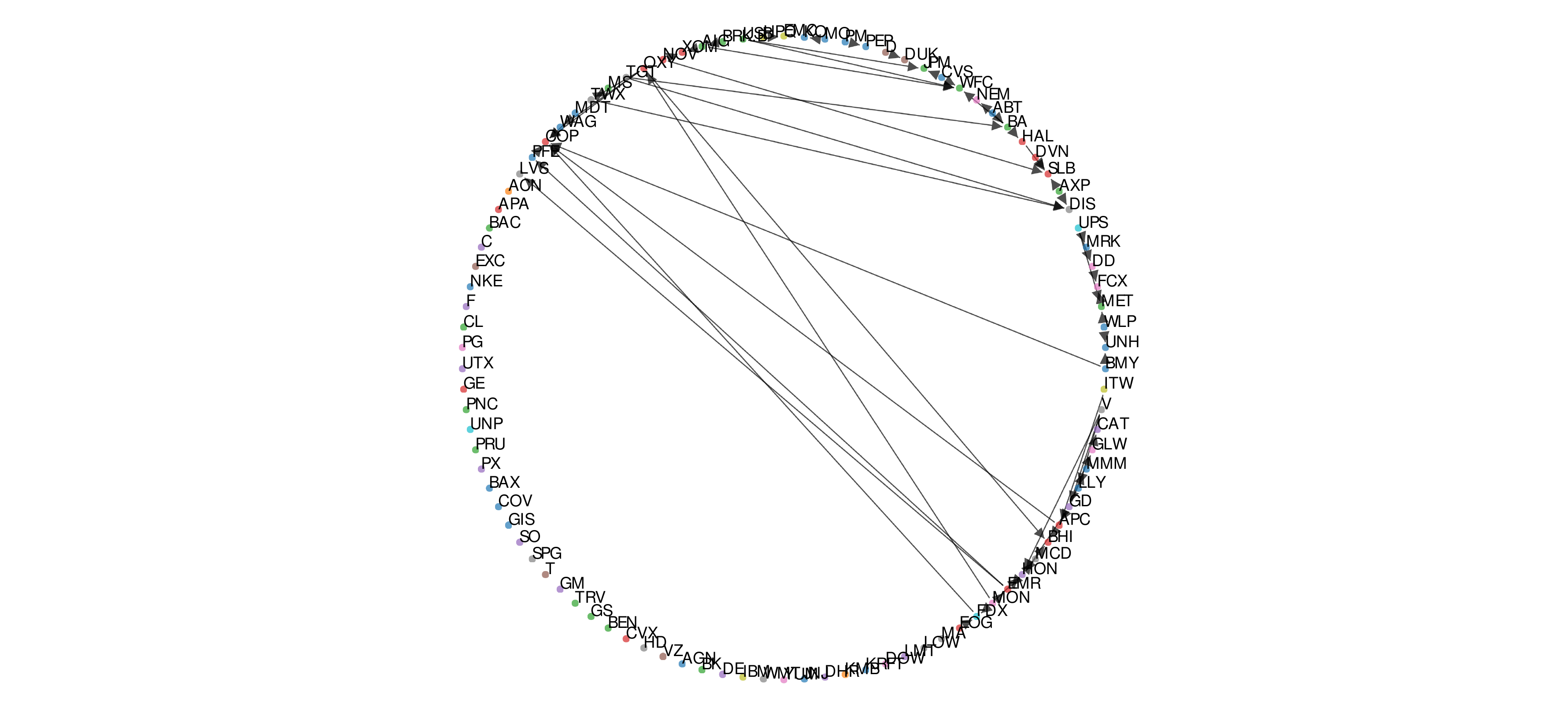}
\caption{MI-based network (intraday, $\lambda=5$)}
\label{fig:nyseid05}
\end{figure}

We have set the p-value to $0.01$ and corrected it using conservative Bonferroni correction. This combined with a choice of an estimator of mutual information which is itself biased downward and thus conservative will give us a very conservative validation of links. One can imagine a much less conservative approach being used. We use the appropriate Gamma distribution for the validation. Moreover while Curme et al. \cite{Curme:2014} set $\lambda$ to be equal to $\tau$ we set $\tau$ to be equal to the interval in the data (1 minute or 1 day) and use $\lambda$ as variable. We find this setup more informative than the one used in \cite{Curme:2014}. As they we impose no topological restraints on the networks.

\begin{figure}[tbh]
\centering
\includegraphics[width=0.5\textwidth]{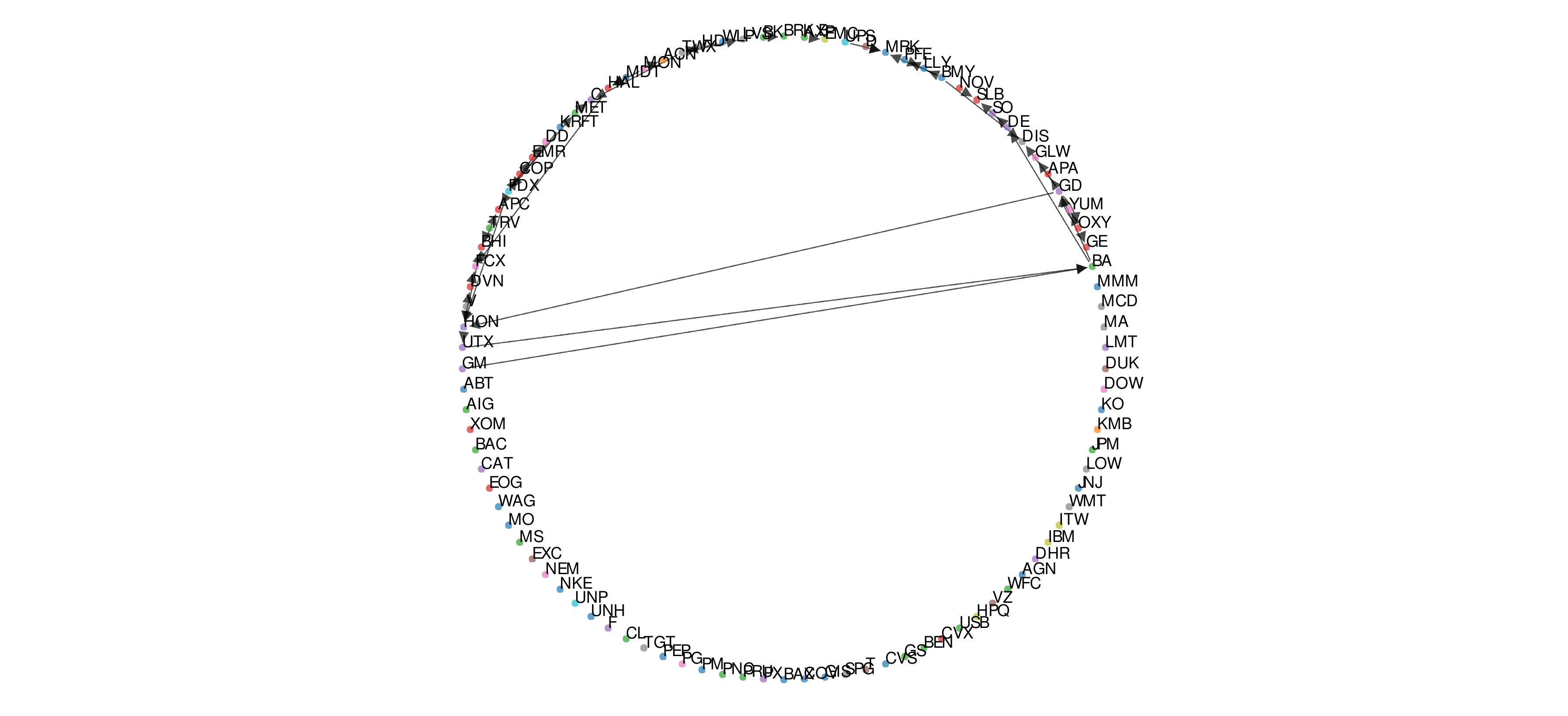}
\caption{MI-based network (intraday, $\lambda=6$)}
\label{fig:nyseid06}
\end{figure}

\begin{figure}[tbh]
\centering
\includegraphics[width=0.5\textwidth]{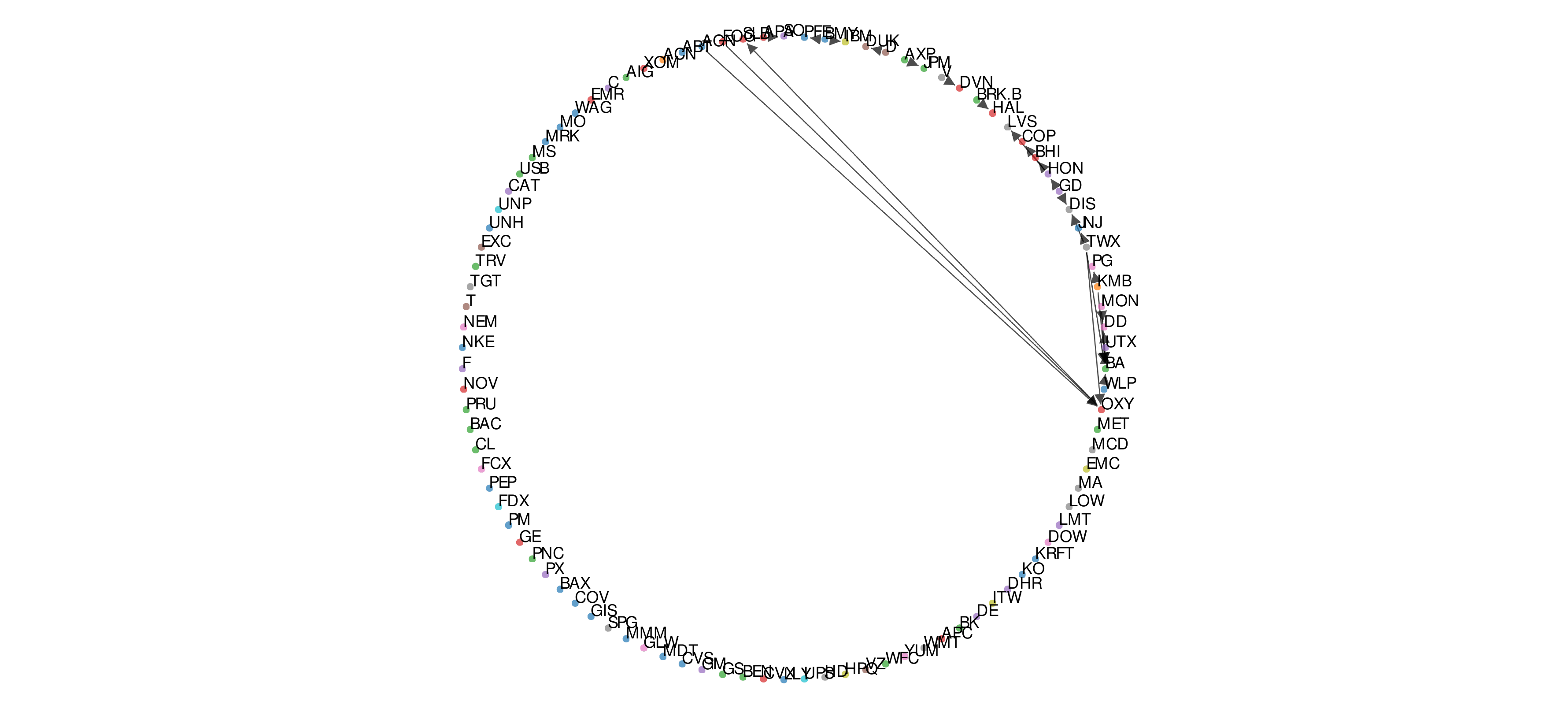}
\caption{MI-based network (intraday, $\lambda=7$)}
\label{fig:nyseid07}
\end{figure}

\begin{figure}[tbh]
\centering
\includegraphics[width=0.5\textwidth]{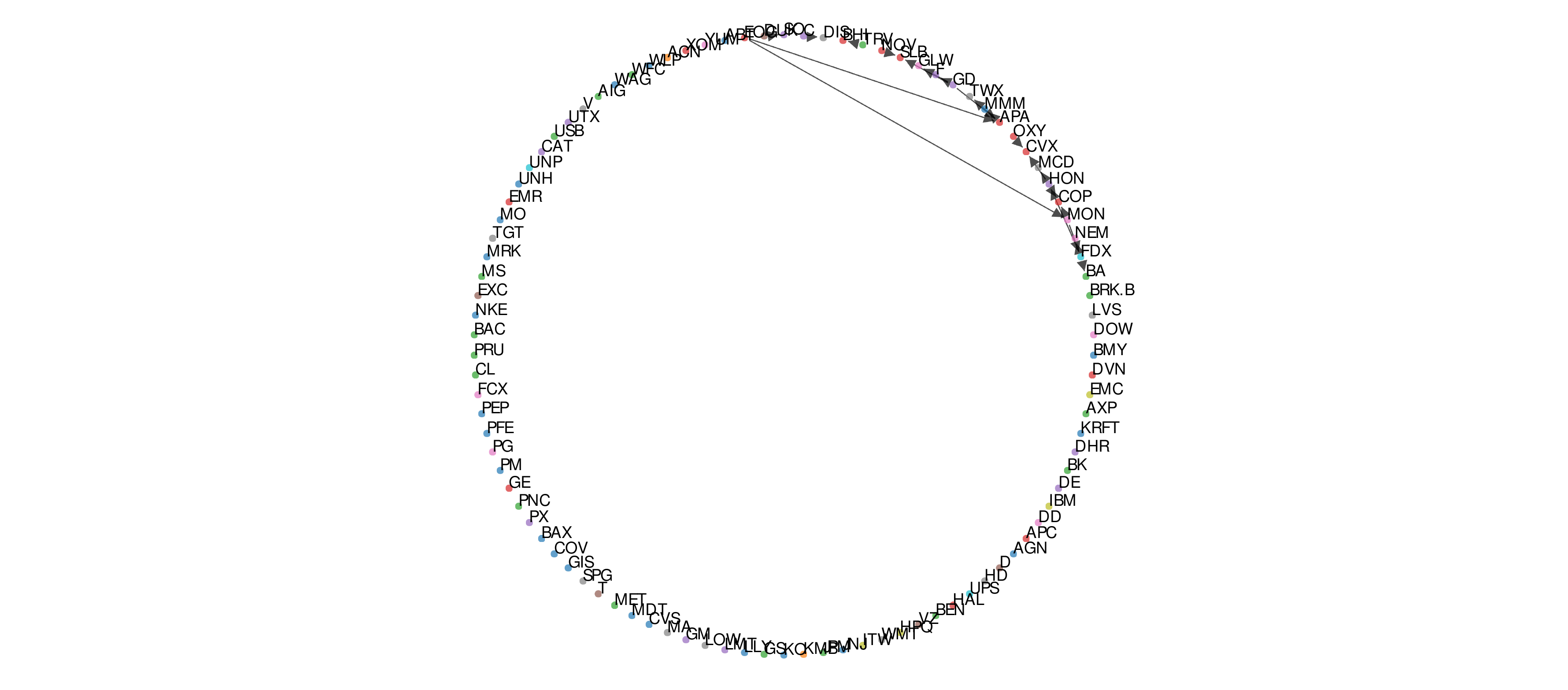}
\caption{MI-based network (intraday, $\lambda=8$)}
\label{fig:nyseid08}
\end{figure}

\begin{figure}[tbh]
\centering
\includegraphics[width=0.5\textwidth]{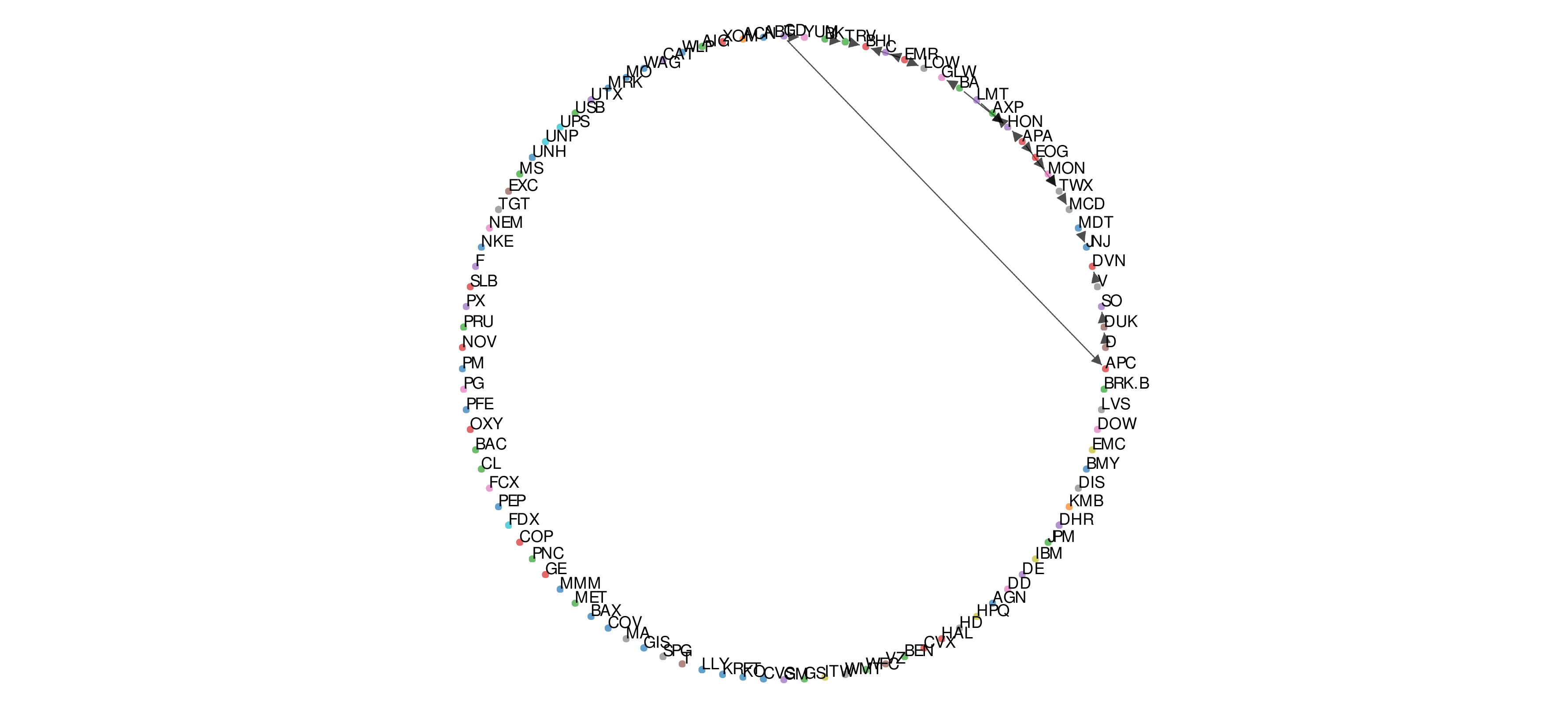}
\caption{MI-based network (intraday, $\lambda=9$)}
\label{fig:nyseid09}
\end{figure}

\begin{figure}[tbh]
\centering
\includegraphics[width=0.5\textwidth]{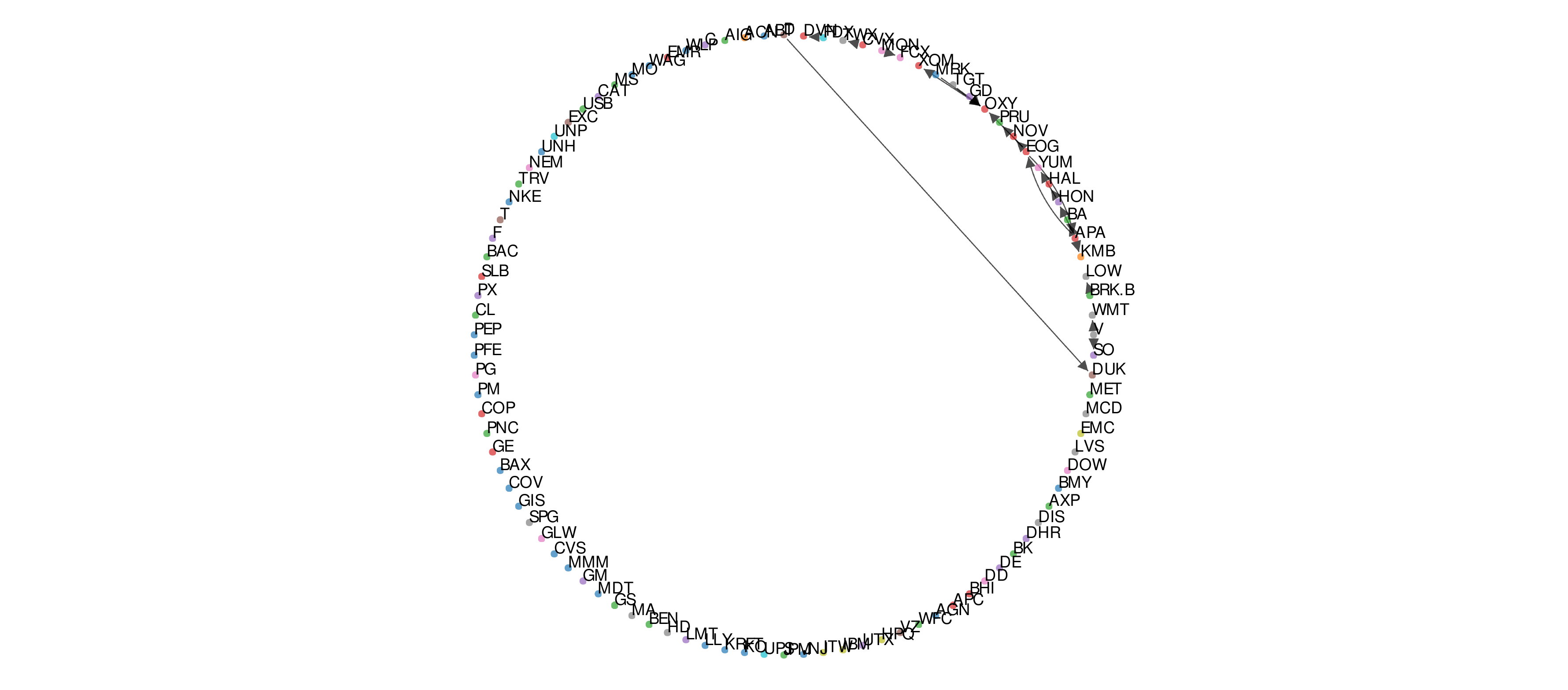}
\caption{MI-based network (intraday, $\lambda=10$)}
\label{fig:nyseid10}
\end{figure}

On Fig.~\ref{fig:nysemigamma} we present the number of validated mutual information-based links for a given shift of $\lambda$ for intraday (1-minute) stock returns. Note that for $\lambda=0$ we create a synchronous network. The networks themselves for different values of $\lambda$ are shown on Figs.~\ref{fig:nyseid00}-\ref{fig:nyseid10}.

\begin{figure}[tbh]
\centering
\includegraphics[width=0.35\textwidth]{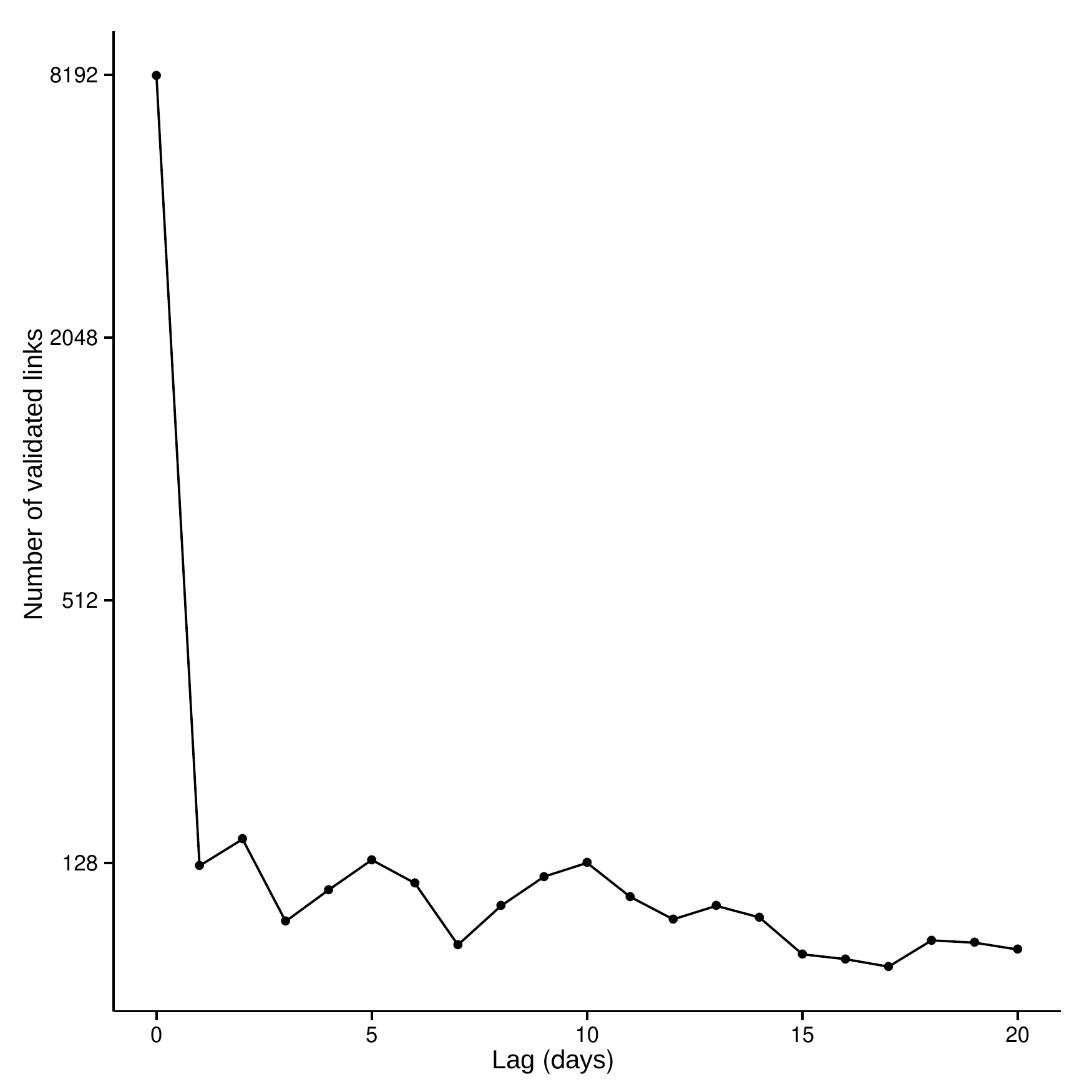}
\caption{Numer of validated links vs lag (daily)}
\label{fig:nysemidaily}
\end{figure}

\begin{figure}[tbh]
\centering
\includegraphics[width=0.4\textwidth]{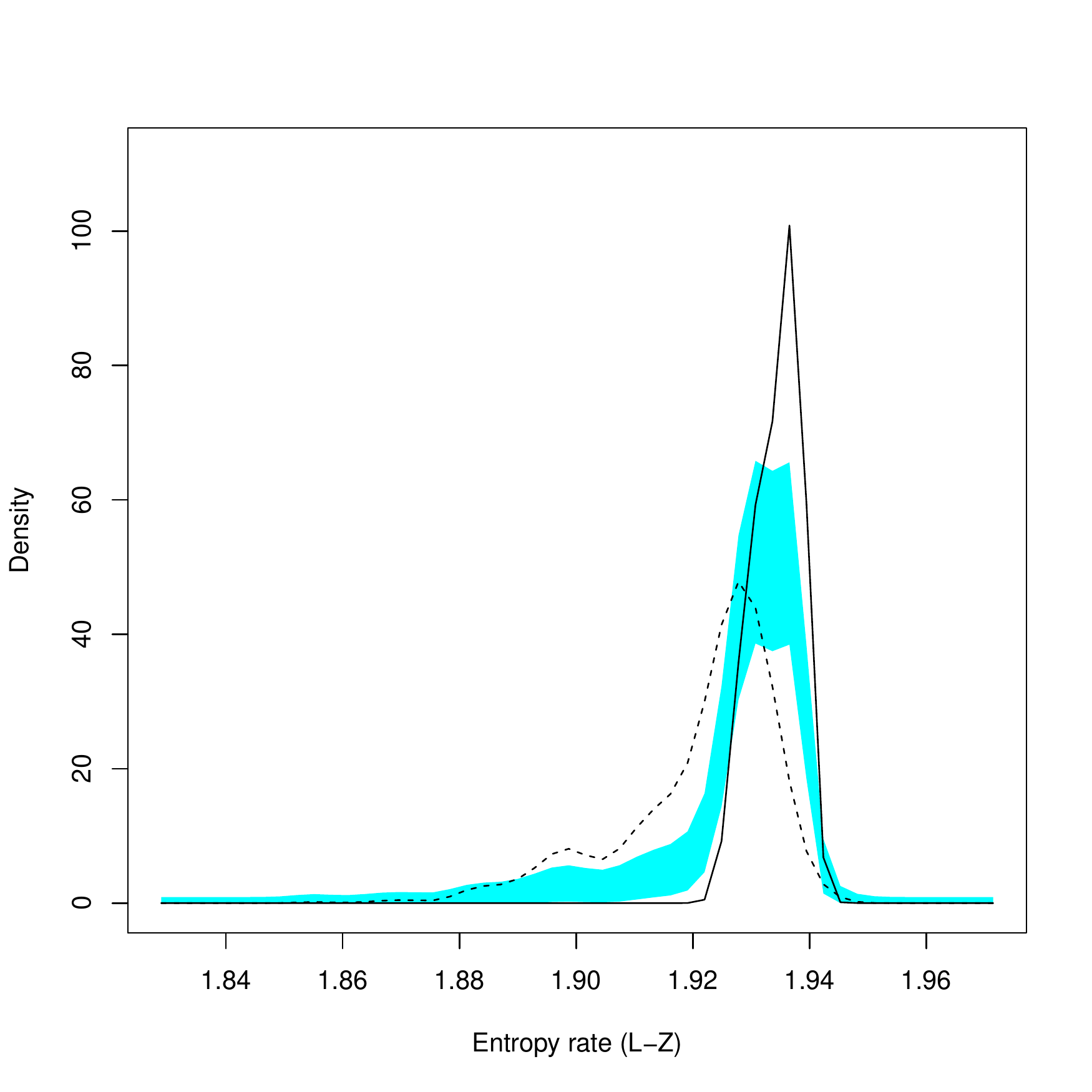}
\caption{Entropy rate -- validated vs non-validated ($\lambda=0$)}
\label{fig:nysedaily00}
\end{figure}

\begin{figure}[tbh]
\centering
\includegraphics[width=0.4\textwidth]{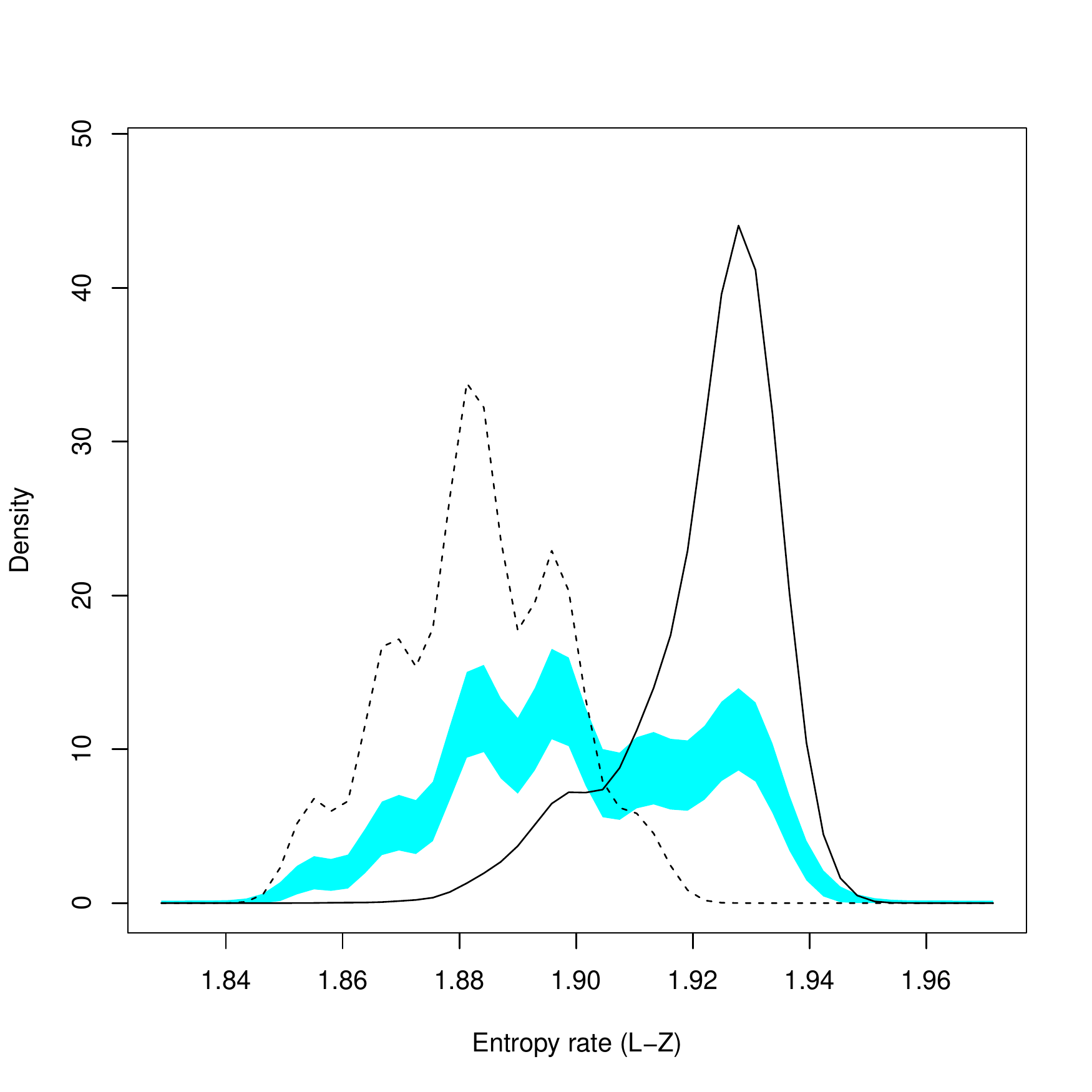}
\caption{Entropy rate -- validated vs non-validated ($\lambda=1$)}
\label{fig:nysedaily01}
\end{figure}

\begin{figure}[tbh]
\centering
\includegraphics[width=0.4\textwidth]{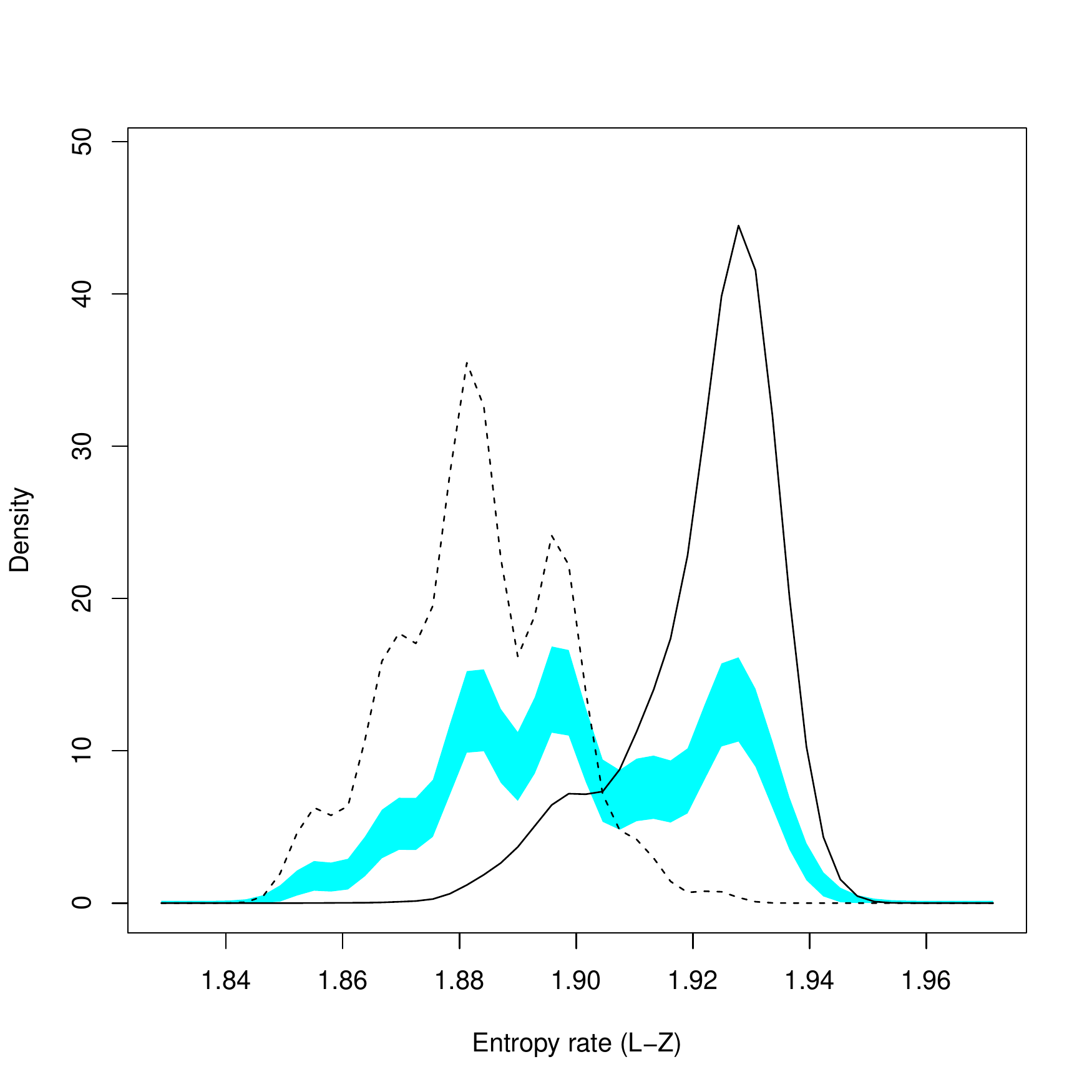}
\caption{Entropy rate -- validated vs non-validated ($\lambda=2$)}
\label{fig:nysedaily02}
\end{figure}

\begin{figure}[tbh]
\centering
\includegraphics[width=0.4\textwidth]{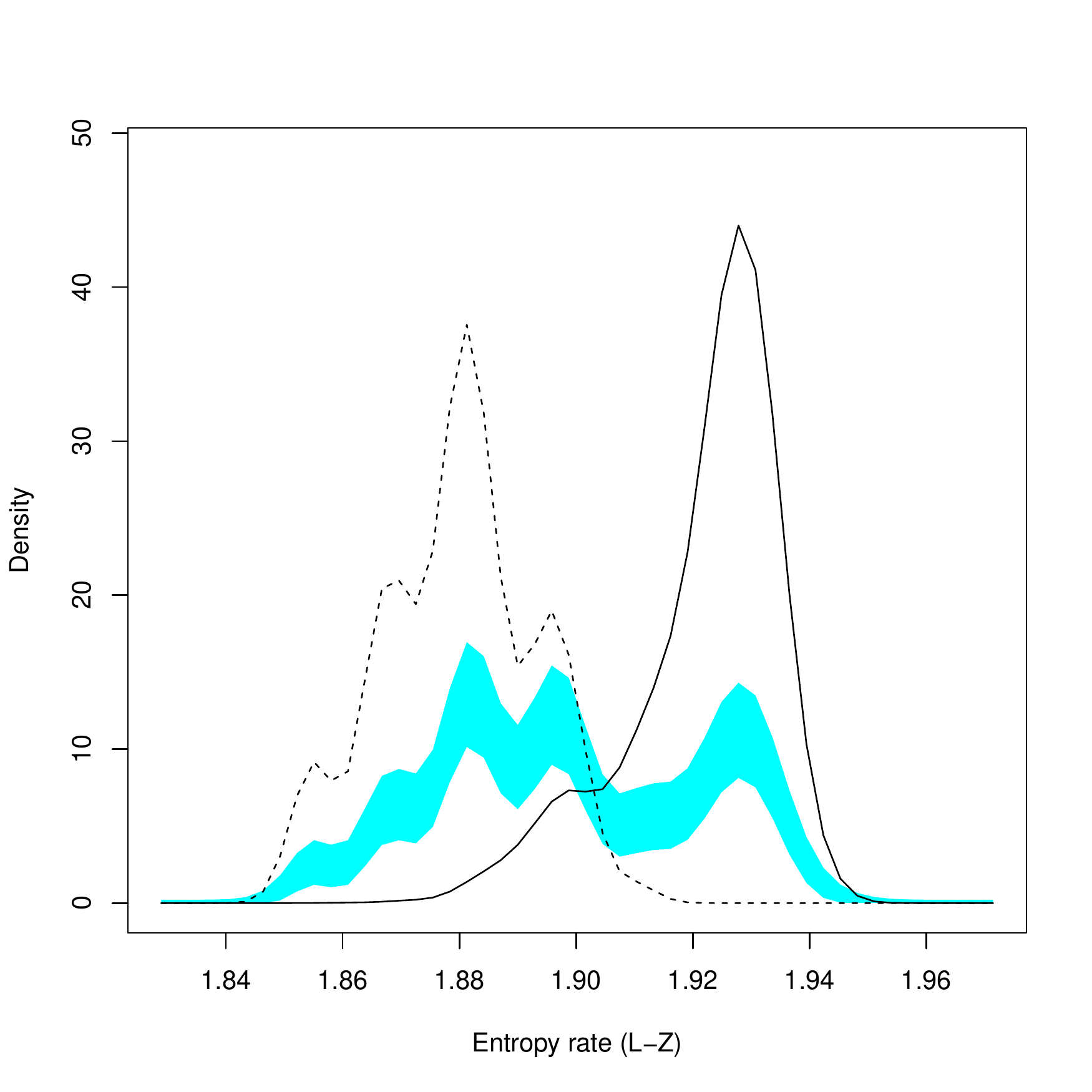}
\caption{Entropy rate -- validated vs non-validated ($\lambda=3$)}
\label{fig:nysedaily03}
\end{figure}

\begin{figure}[tbh]
\centering
\includegraphics[width=0.4\textwidth]{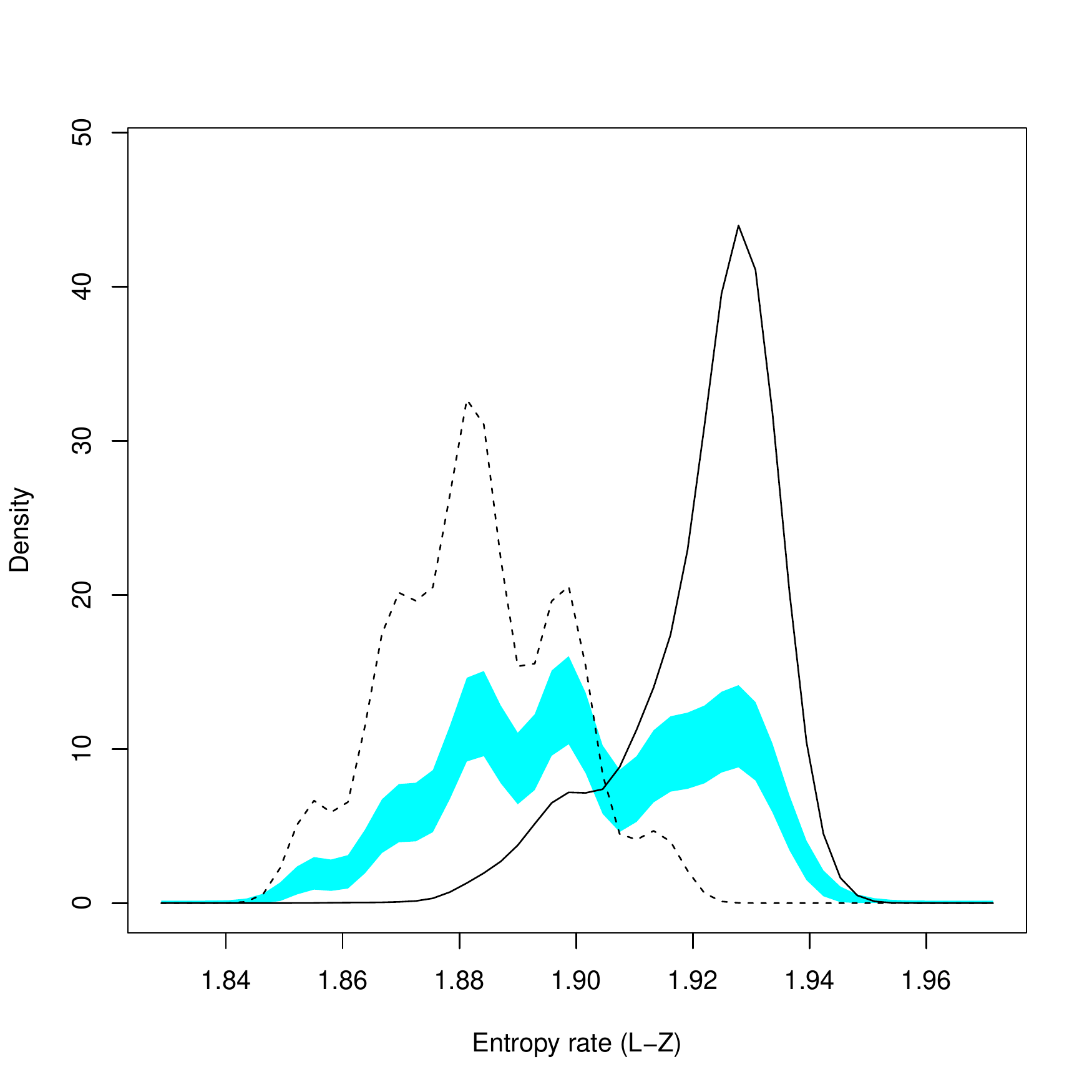}
\caption{Entropy rate -- validated vs non-validated ($\lambda=10$)}
\label{fig:nysedaily10}
\end{figure}

\begin{figure}[tbh]
\centering
\includegraphics[width=0.4\textwidth]{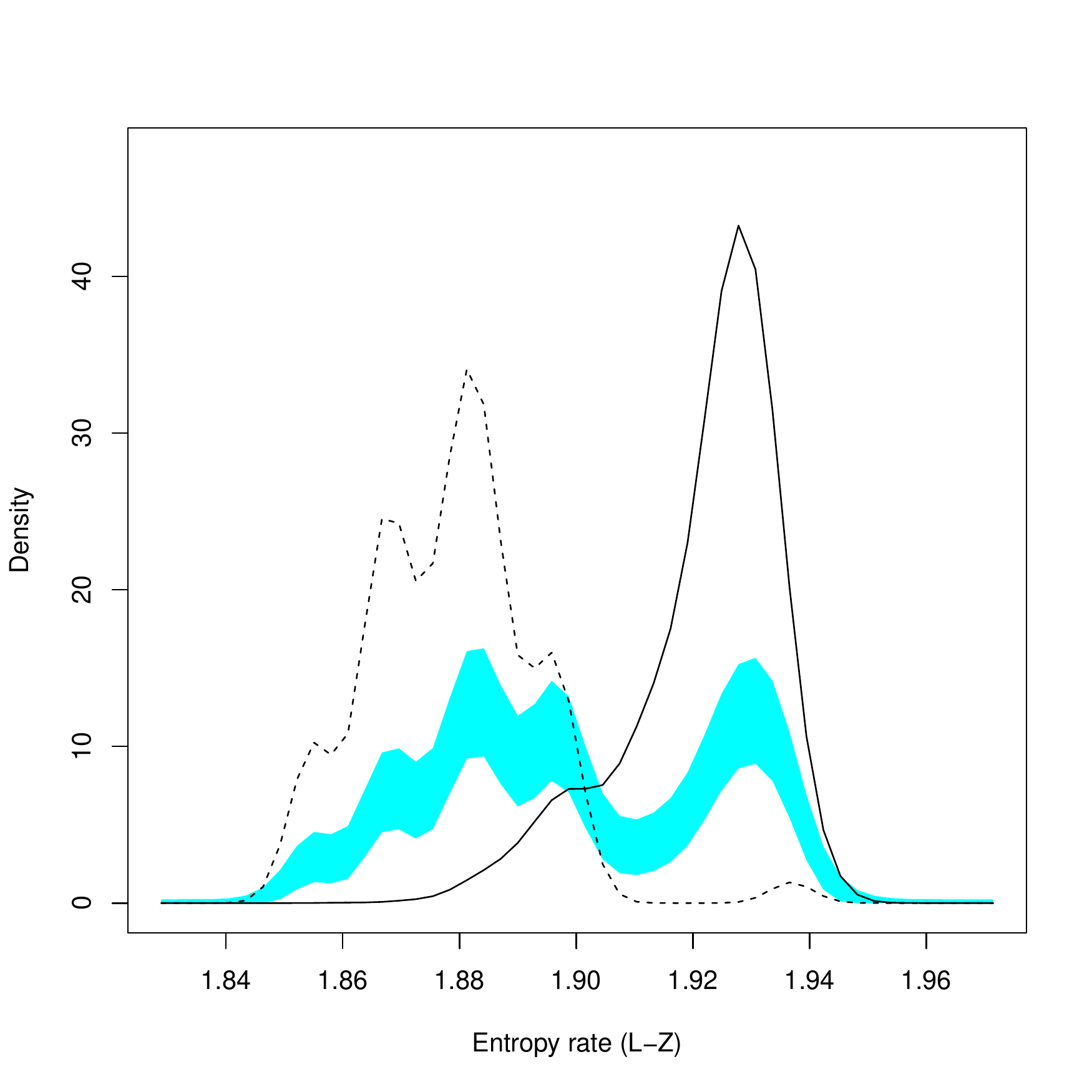}
\caption{Entropy rate -- validated vs non-validated ($\lambda=20$)}
\label{fig:nysedaily20}
\end{figure}

On Fig.~\ref{fig:nysemidaily} we present the number of validated mutual information-based links for a given shift of $\lambda$ for daily stock returns. Note that for $\lambda=0$ we create a synchronous network. The networks themselves for daily data are less illustrative and have been ignored. On Figs.~\ref{fig:nysedaily00}-\ref{fig:nysedaily20} we do show how the validation of links is connected with the entropy rate of the underlying time series (average for the two stocks in a link) for varying values of $\lambda$. We calculate entropy rate using the same data with Lempel-Ziv algorithm, see \cite{Fiedor:2014} for details. The larger the value of entropy rate the more random the price formation process is, in this particular case $2$ being the theoretical maximum for fully random processes and $0$ being the minimum for fully predictable ones.

\section{Discussion}

First, we turn to the intraday analysis. We find that mutual information-based methodology presented in this paper is performing well in validating the lead-lag relationships between financial instruments. Since we use stable $\tau$ and variable $\lambda$ which is different from \cite{Curme:2014} we cannot directly compare the results. They did however briefly comment on the fact that for $\tau$ equal to 15 minutes the number of validated links decreases very quickly with increasing $\lambda$ and is close to $0$ at $\lambda=4$. In our study for $\tau$ equal to 1 minute we find that the the number of validated links decreases much slower and is only irrelevant after $\lambda$ reaches 10 (minutes), as can be seen on Fig.~\ref{fig:nysemigamma}. We thus conclude that the market is quite far from the Efficient Market Hypothesis at such small intervals, which is corroborated not only by \cite{Curme:2014}, but also studies not using network approach \cite{Fiedor:2014}. The mentioned decrease can be easily spotted on Figs.~\ref{fig:nyseid00}-\ref{fig:nyseid10}, which show the Bonferroni networks for $\lambda$ between $0$ and $10$.

Second, we turn into the analysis of daily stock returns. It is often ignored as studies show that daily stock returns are much closer to being random and ruled by EMH than intraday stock returns \cite{Fiedor:2014}. Nonetheless we see on Fig.~\ref{fig:nysemidaily} that while there is an enormous drop of the number of validated links between synchronous ($\lambda=0$) and asynchronous ($\lambda>0$) networks, there is nonetheless a large number of links which are present even at large values of $\lambda$. Curious as to whether these result from statistical uncertainty and noise we compare the predictability of the studied time series between two groups: the validated pairs (dotted lines) and the non-validated pairs (solid lines), as can be seen on Figs.~\ref{fig:nysedaily00}-\ref{fig:nysedaily20} in the form of kernel densities, for the values of $\lambda$ between $0$ and $20$. We find that the stock returns involved in validated pairs are on average significantly more predictable than the ones not involved in validated pairs (the reference band presented is associated with the permutation test for equality). We are therefore inclined to say that these links are not strictly a noise in the data, but present a serious deviation from the Efficient Market Hypothesis in the daily stock returns for certain stocks. Further studies will be required to analyse these relationships thoroughly.

\section{Conclusions}

We have presented a methodology for statistically validating lead-lag relationships between financial instruments which are able to account for non-linear dependencies in the financial markets. We have also applied this methodology on daily and intraday data for NYSE 100 stocks and have founds it to be performing well. While the results for intraday data are not surprising, with the exception of slower than expected decay of the number of validated lead-lag relationships with the increasing lag shift $\lambda$. The results for daily data show that there are statistically validated links which cannot be easily explained as noise in the data, which is surprising and will require further exhaustive studies. Further studies should also be performed to analyse the usefulness and robustness of this methodology on other markets, both geographically (other world markets) and objectively (currency exchange rates, stock indices). A more exhaustive study with varying lag parameters ($\tau$ \& $\lambda$) should also be performed to further understand the deviation of Efficient Market Hypothesis at different time scales.

\bibliographystyle{epj}
\bibliography{prace}
\end{document}